\documentclass[11pt]{article}  

\usepackage{graphicx}
\usepackage{amssymb, amsmath, amsthm}

\newcommand{\QED}{\hfill QED}
\newcommand{\Exp}{\mbox{\rm E}}
\newcommand{\piv}{\mbox{\rm piv}}
\newcommand{\Prob}{\mbox{\rm Pr}}
\newcommand{\Order}{\mbox{\rm O}}
\newtheorem{theorem}{Theorem}
\newtheorem{lemma}[theorem]{Lemma}
\newtheorem{corollary}[theorem]{Corollary}
\newtheorem{assumption}{Assumption}

\begin{document}

\title{Monte Carlo Methods \\
	for Calculating Shapley-Shubik Power Index \\  
	in Weighted Majority Games
\thanks{
A preliminary version of this paper was presented at 
the 21st Japan-Korea Joint Workshop on Algorithms and Computation (WAAC), 
August 26-27, Fukuoka, Japan, 2018.  \\
This work was supported by JSPS KAKENHI Grant Numbers 26285045, 26242027.}
}

\author{
\begin{tabular}{lll@{}ll}
&Tokyo Institute of Technology&& Yuto Ushioda\\
&Tokyo Institute of Technology&& Masato Tanaka\\
&Tokyo Institute of Technology&& Tomomi Matsui
\end{tabular}
}

\date{\today}

\maketitle
\begin{abstract}
This paper addresses Monte Carlo algorithms 
	for calculating the Shapley-Shubik power index 
	in weighted majority games.
First, we analyze a naive Monte Carlo algorithm and
	discuss the required number of samples.
We then propose an efficient Monte Carlo algorithm
	and show that our algorithm reduces 
	the required number of samples as compared 
	to the naive algorithm. \\
keywords:  
Games/Voting, Probability/Applications, 
 Statistics/Sampling,  Monte Carlo algorithm
\end{abstract}

\section{Introduction}

The analysis of power is a central issue in political science. 
In general, it is difficult to define the idea of power
	even in restricted classes 
 	of the voting rules commonly considered  
	by political scientists.
The use of game theory to study the distribution
	of power in voting systems can be traced back 
	to the invention of ``simple games'' by
	von Neumann and Oskar Morgenstern 
\cite{NEUMANN1944}.
A simple game is an abstraction of the
	constitutional political machinery for voting.

In 1954, Shapley and Shubik~\cite{SHAPLEY1954}
	proposed the specialization 
	of the Shapley value~\cite{SHAPLEY1953}  
	to assess the a priori measure 
	of power of each player in a simple game.
Since then, the Shapley-Shubik power index (S-S index)
	has become widely known 
	as a mathematical tools for measuring the relative power 
	of the players in a simple game.

In this paper, 
	we consider a special class of simple games, 
	called {\em weighted majority games},
	which constitute a familiar example of voting systems.
Let $N$ be a set of players. 
Each player $i \in N$ has a positive integer voting weight $w_i$
	as the number of votes or weight of the player.
The quota needed for a coalition to win 
	is a positive integer $q$.
A coalition $N' \subseteq N$ is a {\em winning coalition}, 
	if $\sum_{i \in N'} w_i \geq q$ holds; 
	otherwise, it is a {\em losing coalition}. 

The difficulty involved in calculating the S-S index 
	in weighted majority games
	is described in \cite{GAREY1979}
	without proof (see p.~280, problem [MS8]).
Deng and Papadimitriou~\cite{DENG1994} showed 
	the problem of computing the S-S index 
	in weighted majority games to be \mbox{$\#$P-complete}.
Prasad and Kelly~\cite{PRASAD1990} 
	proved the \mbox{NP-completeness}
	of the problem of verifying the positivity 
	of a given player's \mbox{S-S index} in weighted majority games. 
The problem of verifying the asymmetricity
	of a given pair of players was also shown 
	to be \mbox{NP-complete}~\cite{MATSUI2001}.
It is known that 
	even approximating the S-S index 	
	within a constant factor 
	is intractable unless $\mbox{P}=\mbox{NP}$~\cite{ELKIND2007}.

There are variations of methods 
	for calculating the S-S index.
These include algorithms based on	
	the Monte Carlo method~\cite{MANN1960,MATSUI2000,FATIMA2008,CASTRO2009,BACHRACH2010,CASTRO2017},
	multilinear extensions~\cite{OWEN1972,LEECH2003},
	dynamic programming~\cite{BRAMS1975,LUCAS1983,MANN1962,MATSUI2000,UNO2012},
	generating functions~\cite{BILBAO2000},
	binary decision diagrams~\cite{BOLUS2011},
	the Karnaugh map~\cite{RUSHDI2017},
	relation algebra~\cite{BERGHAMMER2011}, or
	the enumeration technique~\cite{KLINZ2005}.
A survey of algorithms 
	for calculating power indices 
	in weighted voting games is presented in~\cite{MATSUI2000}.

This paper addresses Monte Carlo algorithms 
	for calculating the S-S index 
	in weighted majority games.
In the following section, 
	we describe the notations and definitions
	used in this paper.
In Section~\ref{naiveAlgorithm}, 
	we analyze a naive Monte Carlo algorithm
	(Algorithm~A1) and
	extend some results obtained 
	in the study reported in~\cite{BACHRACH2010}.
In Section~\ref{efficientAlgorithm}, 
	we propose an efficient Monte Carlo algorithm
	(Algorithm~A2)
	and show that our algorithm reduces 
	the required number of samples as compared 
	to the naive algorithm.  
Table~\ref{numberofsamples} summarizes the results 
	of this study,
	where  
	$(\varphi_1,\varphi_2,\ldots ,\varphi_n)$ denotes 
	the S-S index and
	$(\varphi^{\mbox{A}}_1,\varphi^{\mbox{A}}_2,
			\ldots ,\varphi^{\mbox{A}}_n)$	
	denotes the estimator obtained by Algorithm~A1 or A2.

\begin{table}[htb]
\caption{Required Number of Samples.} \label{numberofsamples}
\begin{center}
\begin{tabular}{|l|c|c|}
\hline
& \multicolumn{2}{|c|}{Required number of samples}\\ 
\cline{2-3} 
Property & Algorithm~A1 & Algorithm~A2 \\
& (naive algorithm) &  (our algorithm) \\
\hline
&& \\[-2ex]
$ \displaystyle
\Prob 
	\left[ 
		\left|
			\varphi^{\mbox{A}}_i-\varphi_i
		\right| < \varepsilon  
	\right] \geq 1-\delta 
$ 
&	$\displaystyle \frac{\ln 2+ \ln (1/\delta)}{2\varepsilon^2}$
& $\displaystyle 
		\frac{\ln 2+ \ln (1/\delta)}{2\varepsilon^2} 
		\left( \frac{1}{i^2} \right)$ \\
&
\multicolumn{1}{r|}{(Bachrach et al.~\cite{BACHRACH2010})}
& (assume $w_1\geq  \cdots \geq w_n$)\\
\hline
&& \\[-2ex]
$ \displaystyle 
 \Prob 
	\left[
		\forall i \in N, 
		\left|			
			\varphi^{\mbox{A}}_i-\varphi_i
		\right| < \varepsilon 
	\right] \geq 1-\delta
$ 
& $\displaystyle \frac{\ln 2+\ln (1/\delta)+\ln n}{2\varepsilon^2}$
& $\displaystyle 
		\frac{\ln 2 +\ln (1/\delta)+\ln 1.129}{2\varepsilon^2}$
\\[2ex]
\hline
&& \\[-2ex]
$ \displaystyle 
	\Prob
	\left[
		\frac{1}{2}\sum_{i \in N} 
		\left|
			\varphi^{\mbox{A}}_i-\varphi_i
		\right| < \varepsilon
	\right] \geq 1- \delta.
$ 
& $\displaystyle \frac{n \ln 2+ \ln (1/\delta)}{2\varepsilon^2}$
& $\displaystyle 
		\frac{n'' \ln 2+ \ln (1/\delta)}{2\varepsilon^2}$
\\[2ex]
\hline
\end{tabular}
\end{center}
An integer $n''$ denotes the size of a maximal player subset 
	with mutually different weights.
\end{table}

\section{Notations and Definitions}\label{notation}

In this paper, 
	we consider a special class of cooperative games 
	called {\em weighted majority games}. 
Let $N = \{1, 2,\ldots , n\}$ be a set of {\em players}. 
A subset of players is called a {\em coalition}.
A weighted majority game $G$ is defined 
	by a sequence of positive integers
	$G=[q;w_1,w_2,\ldots ,w_n]$,
	where we may think of $w_i$ as the number of votes
	or the weight of player $i$ and $q$ as the quota 
	needed for a coalition to win. 
In this paper, we assume that $0< q \leq w_1+w_2 + \cdots+ w_n$. 

A coalition $S \subseteq N$ is called a {\em winning coalition}
	when the inequality $q \leq \sum_{i \in S}w_i $ holds. 
The inequality $q \leq w_1+w_2 + \cdots+ w_n$ implies that
	$N$ is a winning coalition.
A coalition $S$ is called a {\em losing coalition} 
	if $S$ is not winning. 
We define that an empty set is a losing coalition.

Let $\pi: \{1,2,\ldots ,n\} \rightarrow N$ 
	be a permutation defined on the set of players $N$,
	which provides a sequence of players 
	$(\pi(1), \pi(2), \ldots ,\pi(n))$.
We denote the set of all the permutations by $\Pi_N$. 
We say that the player $\pi(i) \in N$ is the {\em  pivot} of 
	the permutation $\pi \in \Pi_N$,
	if $\{\pi(1), \pi(2), \ldots , \pi(i-1)\}$ is a losing coalition
	and $\{\pi(1), \pi(2), \ldots , \pi(i-1), \pi(i)\}$
	is a winning coalition. 
For any permutation $\pi \in \Pi_N$, 
	$\piv (\pi)\in N$ denotes the pivot of $\pi$.
For each player $i \in N$, 
	we define
	$\Pi_i=\{\pi \in \Pi_N \mid \piv(\pi)=i \}$.
Obviously, $\{ \Pi_1, \Pi_2, \ldots ,\Pi_n\}$
	becomes a partition of $\Pi_N$. 
The S-S index
	 of player $i$, denoted by $\varphi_i$,
	 is defined by $|\Pi_i|/n!$.
Clearly, we have that 
	$0 \leq \varphi_i \leq 1 \; (\forall i \in N)$
	and $\sum_{i \in N} \varphi_i=1$.

\begin{assumption}\label{monotone}
	 The set of players is arranged to satisfy  
	$w_1\geq w_2 \geq \cdots \geq w_n$.
\end{assumption}

\noindent
Clearly, this assumption implies that
	$\varphi_1 \geq \varphi_2 \geq \cdots \geq \varphi_n$.
	
\section{Naive Algorithm and its Analysis}
\label{naiveAlgorithm}

In this section, 
	we describe a naive Monte Carlo algorithm and
	analyze its theoretical performance.

\noindent
\begin{description}
\item[\underline{Algorithm A1}]
\item[Step~0:]
	Set $m:=1$, $\varphi'_i:=0 \;\; (\forall i \in N)$.
\item[Step~1:] 
	Choose 
	$\pi \in \Pi_N$ 	uniformly at random. \\
	\indent	Put (the random variable) $I^{(m)}:=\piv (\pi)$. 
	Update $\varphi'_{I^{(m)}}:=\varphi'_{I^{(m)}}+1$. 
\item[Step~2:] If $m=M$, then output  
	$\varphi'_i/M \;\; (\forall i \in N)$ and stop.\\
	\indent	Else, update $m:=m+1$ and go to Step~1.
\end{description}

\medskip 

For each permutation $\pi \in \Pi_N$,
	we can find the pivot $\piv (\pi) \in N$ 
	in $\Order (n)$ time.
Thus, the time complexity of Algorithm~A1
	is bounded by $\Order (M(\tau (n)+n))$
	where $\tau(n)$ denotes the computational effort
	required for random generation of a permutation.

We denote the vector (of random variables) 
	obtained by Algorithm~A1
	by 	
	$(\varphi^{\mbox{A1}}_1, \varphi^{\mbox{A1}}_2,\ldots ,
		\varphi^{\mbox{A1}}_n)$.
The following theorem is obvious.

\begin{theorem}
For each player $i \in N$, 
	$\Exp \left[ \varphi^{\mbox{\rm A1}}_i \right]=\varphi_i$.
\end{theorem}	

The following theorem provides the number of 
	samples required in Algorithm~A1.

\begin{theorem}\label{sampleA1}
For any $\varepsilon >0$ and  $0< \delta <1$,
	we have the following.

{\rm (1)~\cite{BACHRACH2010}} 
If we set 
	$\displaystyle M \geq \frac{\ln 2+\ln (1/\delta)}{2\varepsilon^2}$,
	then each player $i \in N$ satisfies that
\[
\Prob 
	\left[ 
		\left|
			\varphi^{\mbox{\rm A1}}_i-\varphi_i
		\right| < \varepsilon  
	\right] \geq 1-\delta.
\] 

{\rm (2)}
 If we set 	
	$\displaystyle M\geq \frac{\ln 2+ \ln (1/\delta)+\ln n}{2\varepsilon^2}$,
	then 
\[
 \Prob 
	\left[
		\forall i \in N, 
		\left|			
			\varphi^{\mbox{\rm A1}}_i-\varphi_i
		\right| < \varepsilon 
	\right] \geq 1-\delta.
\]

{\rm (3)}
 If we set 	
	$\displaystyle M\geq \frac{n \ln 2+ \ln (1/\delta)}{2\varepsilon^2}$,
	then 
\[
	\Prob
	\left[
		\frac{1}{2}\sum_{i \in N} 
		\left|
			\varphi^{\mbox{\rm A1}}_i-\varphi_i
		\right| < \varepsilon
	\right] \geq 1- \delta.
\]
\end{theorem}

\noindent
The distance measure 	
	$\frac{1}{2}\sum_{i \in N} 
		\left|
			\varphi^{\mbox{A1}}_i-\varphi_i
		\right|
	$ appearing in~(3) is called 
	the {\em total variation distance}.

\noindent
Proof.  
Let us introduce random variables 
	$X^{(m)}_i \; (\forall m \in \{1,2,\ldots ,M\}, \forall i \in N)$
	in Step~1 of Algorithm~A1 defined by
\[
		X^{(m)}_i=
		\left\{
			\begin{array}{ll}
				1 & (\mbox{if } i= I^{(m)}),\\
				0		& (\mbox{otherwise}).
			\end{array}
		\right.
\]
It is obvious that for each player $i \in N$, 
	$\{ X^{(1)}_i, X^{(2)}_i, \ldots , X^{(M)}_i\}$ 
	is a Bernoulli process satisfying
	$\varphi^{\mbox{A1}}_i=\sum_{m=1}^M X^{(m)}_i/M$,
 	$ \Exp \left[ \varphi^{\mbox{A1}}_i \right]
	=\Exp \left[ X^{(m)}_i \right]=\varphi_i$ 
	$(\forall m \in \{1,2,\ldots ,M\})$.
Hoeffding's inequality~\cite{HOEFFDING1963}
	implies that 
	each player $i \in N$ satisfies
\begin{eqnarray*}
\Prob \left[ \left| \varphi^{\mbox{A1}}_i
	- \varphi_i \right|
	\geq \varepsilon \right]
&\leq 
& 2 \exp \left(
	- \frac{2M^2\varepsilon^2}{\sum_{m=1}^M (1-0)^2}
	\right)
= 2 \exp (-2M\varepsilon^2).
\end{eqnarray*}

\noindent
(1) If we set 
	$\displaystyle M \geq \frac{\ln (2/\delta)}{2\varepsilon^2}$,
	then 
\[
\Prob \left[ \left| \varphi^{\mbox{A1}}_i
	- \varphi_i \right|
	< \varepsilon \right]\geq 
	1-2 \exp 
	\left(
		-2 \frac{\ln (2/\delta)}{2\varepsilon^2}
		\varepsilon^2
	\right)=1-\delta.
\]

\noindent
(2) If we set 
	$\displaystyle M\geq \frac{\ln (2n/\delta)}{2\varepsilon^2}$,
	then we have that
\begin{eqnarray*}
\lefteqn{
\Prob \left[
				\forall i \in N, 
					\left| 
						\varphi^{\mbox{A1}}_i	- \varphi_i 
					\right|
					< \varepsilon 
			\right]
 = 1- \Prob 
			\left[
				\exists i \in N, 
					\left| 
						\varphi^{\mbox{A1}}_i	- \varphi_i 
					\right|
					\geq \varepsilon 
			\right]}\\
& \geq 
& 1-\sum_{i \in N} 
	\Prob 
			\left[
					\left| 
						\varphi^{\mbox{A1}}_i	- \varphi_i 
					\right|
					\geq \varepsilon 
			\right]
\geq 1- \sum_{i=1}^n  2 \exp (-2M\varepsilon^2)\\
& \geq 
& 1- \sum_{i=1}^n  2 \exp 
	\left( 
		-2  \frac{\ln (2n/\delta)}{2\varepsilon^2} 
     \varepsilon^2 
	\right) 
= 1-\sum_{i=1}^n \frac{\delta}{n}=1-\delta.  
\end{eqnarray*}

\noindent
(3) 
Obviously, the vector of random variables
\[
	(M\varphi^{\mbox{A1}}_1, M\varphi^{\mbox{A1}}_2, \cdots,
    M\varphi^{\mbox{A1}}_n)
	=\left( 
		\sum_{m=1}^M X^{(m)}_1, \sum_{m=1}^M X^{(m)}_2, \cdots,
		\sum_{m=1}^M X^{(m)}_n
	\right)
\]
	is multinomially distributed with parameters
	$M$ and $(\varphi_1, \varphi_2, \cdots, \varphi_n)$.
Then, the Bretagnolle-Huber-Carol inequality~\cite{VAART1989}
	(Theorem~\ref{BHC} in Appendix)
	implies that
\begin{eqnarray*}
	\Prob
	\left[
		\frac{1}{2}\sum_{i \in N} 
		\left|
			\varphi^{\mbox{A1}}_i-\varphi_i
		\right| \geq \varepsilon
	\right] 
&=& \Prob
	\left[
		\sum_{i \in N} 
		\left|
			M\varphi^{\mbox{A1}}_i - M\varphi_i
		\right| \geq 2M\varepsilon 
	\right]
\leq 2^n \exp 
	\left( -2M \varepsilon^2 \right) \\
&\leq & 2^n \exp 
	\left( -2
		\left(
			\frac{\ln (2^n/\delta)}{2 \varepsilon^2}
		\right)	
 		\varepsilon^2 
	\right) = \delta, \\
\end{eqnarray*}
and thus, we have the desired result.
\qed

\smallskip

\section{Our Algorithm}\label{efficientAlgorithm}

In this section, we propose a new algorithm
	based on the hierarchical structure 
	of the partition $\{\Pi_1, \Pi_2, \ldots ,\Pi_n\}$.
First, we introduce a map 
	$f_i : \Pi_i \rightarrow \Pi_N$
	for each $i \in N\setminus \{1\}$.
For any $\pi \in \Pi_i$, 
	$f_i(\pi)$ denotes a permutation 
	obtained by swapping 
	the positions of players $i$ and $i-1$
	in the permutation $(\pi(1), \pi(2), \ldots, \pi(n))$.
Because $w_{i-1}\geq w_i$ (Assumption~\ref{monotone}),
	it is easy to show that 
	the pivot of $f_i(\pi)$ becomes the player $i-1$.
The definition of $f_i$ directly implies that
	$\forall \{\pi,\pi'\} \subseteq \Pi_i$,
	if $\pi \neq \pi'$, then $f_i(\pi) \neq f_i(\pi')$.
Thus, we have the following.
\begin{lemma}\label{injection}
For any $i \in N\setminus \{1\}$,
	the map $f_i:\Pi_i \rightarrow \Pi_{i-1}$
	is injective.
\end{lemma}

\begin{figure}[h]
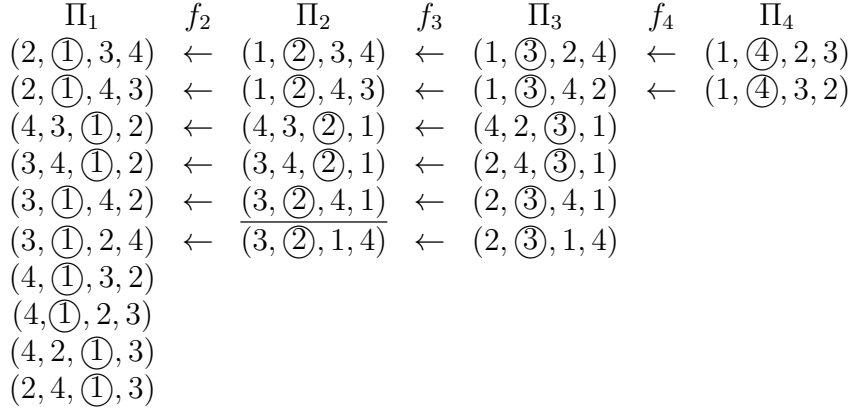

\[
\large 
\begin{array}{ccccccc}
\Pi_1 & f_2 & \Pi_2 & f_3 & \Pi_3 & f_4 & \Pi_4 \\
(2,\mbox{\textcircled{1}},3,4)& \leftarrow &
(1,\mbox{\textcircled{2}},3,4)& \leftarrow &
(1,\mbox{\textcircled{3}},2,4)& \leftarrow &
(1,\mbox{\textcircled{4}},2,3) \\
(2,\mbox{\textcircled{1}},4,3)& \leftarrow &
(1,\mbox{\textcircled{2}},4,3)& \leftarrow &
(1,\mbox{\textcircled{3}},4,2)& \leftarrow &
(1,\mbox{\textcircled{4}},3,2) \\ 
(4,3,\mbox{\textcircled{1}},2)& \leftarrow &
(4,3,\mbox{\textcircled{2}},1)& \leftarrow &
(4,2,\mbox{\textcircled{3}},1)\\
(3,4,\mbox{\textcircled{1}},2)& \leftarrow &
(3,4,\mbox{\textcircled{2}},1)& \leftarrow &
(2,4,\mbox{\textcircled{3}},1)\\
(3,\mbox{\textcircled{1}},4,2)& \leftarrow &
\underline{(3,\mbox{\textcircled{2}},4,1)}& \leftarrow &
(2,\mbox{\textcircled{3}},4,1)\\
(3,\mbox{\textcircled{1}},2,4)& \leftarrow &
(3,\mbox{\textcircled{2}},1,4)& \leftarrow &
(2,\mbox{\textcircled{3}},1,4)\\
(4,\mbox{\textcircled{1}},3,2)\\
(4\mbox{,\textcircled{1}},2,3)\\
(4,2,\mbox{\textcircled{1}},3)\\
(2,4,\mbox{\textcircled{1}},3)\\
\end{array}
%
%
%
\]
\begin{center}
\caption{
Injective maps $f_2,f_3,f_4$ induced by $G=[50; 40, 30, 20, 10]$. 
   Circled number (player) denotes the pivot player.}  \label{fig:injection}
\end{center}
\end{figure}

\noindent
Figure~\ref{fig:injection} shows injective maps 
	$f_2,f_3,f_4$ induced by $G=[50; 40, 30, 20, 10]$. 

When an ordered pair of permutations $(\pi, \pi')$ 
	satisfies the conditions 
	that $\pi \in \Pi_i$, $\pi' \in \Pi_j$, $i \leq j$, and
	$\pi=f_{i-1}\circ \cdots \circ f_{j-1} \circ f_j(\pi')$,
	we say that $\pi'$ is an {\em ancestor} of $\pi$.
Here, we note that $\pi$ is always an ancestor of $\pi$ itself.
Lemma~\ref{injection} implies that
	every permutation 
	$\pi \in \Pi_N$
	has a unique ancestor, called the {\em originator},
	$\pi'\in \Pi_j$ 
	satisfying that 
	either  $j=n$ or 
	its inverse image $f^{-1}_{j+1}(\pi')=\emptyset$.
For each permutation $\pi \in \Pi_N$,
	${\rm org} (\pi)\in N$ 
	denotes the pivot of the originator of $\pi$;
	i.e.,  $\Pi_{{\rm org}(\pi)}$ includes the originator of $\pi$.

Now, we describe our algorithm.

\noindent
\begin{description}
\item[\underline{Algorithm A2}]
\item[Step~0:]
	Set $m:=1$, $\varphi'_i:=0 \;\; (\forall i \in N)$.
\item[Step~1:] Choose 
	$\pi \in \Pi_N$ 	uniformly at random. 
	Put the random variable $L^{(m)}:={\rm org} (\pi)$. \\
	\indent Update 
	$\varphi'_i:=
		\left\{ \begin{array}{ll}
			\varphi'_i+1/L^{(m)} 
				& (\mbox{if } 1 \leq  i \leq L^{(m)}),\\
			\varphi'_i     
				& (\mbox{if } L^{(m)} < i).
		\end{array} \right.$ 
\item[Step~2:] If $m=M$, then output  
	$\varphi'_i/M \;\; (\forall i \in N)$ and stop.\\
	\indent Else, update $m:=m+1$ and go to Step~1.
\end{description}

\noindent
In the example shown in Figure~\ref{fig:injection}, 
if we choose $\pi=(3,\mbox{\textcircled{2}},4,1)$
at Step~1 of Algorithm~A2,
then ${\rm org} (\pi)=3$ and 
Algorithm~A2 updates 
\[
(\varphi'_1, \varphi'_2, \varphi'_3, \varphi'_4)
:=(\varphi'_1+(1/3), \varphi'_2+(1/3), \varphi'_3+(1/3), \varphi'_4).
\]

For each permutation $\pi \in \Pi_N$,
	we can find the originator ${\rm org} (\pi) \in N$ 
	in $\Order (n)$ time.
Thus, the time complexity of Algorithm~A2
	is also bounded by $\Order (M(\tau (n)+n))$
	where $\tau(n)$ denotes the computational effort
	required for random generation of a permutation.

We denote the vector (of random variables)
	obtained by Algorithm~A2 by 	
	$(\varphi^{\mbox{A2}}_1, \varphi^{\mbox{A2}}_2,\ldots ,
		\varphi^{\mbox{A2}}_n)$.
The following theorem is obvious.

\begin{theorem}
{\rm (1)}
 For each player $i \in N$, 
	$\Exp \left[ \varphi^{\mbox{\rm A2}}_i \right]=\varphi_i$.

\noindent
{\rm (2)} 
For each pair of players $\{i,j\} \subseteq N$, 
	if $\varphi_i > \varphi_j$, 
	then
	 $\varphi^{\mbox{\rm A2}}_i \geq \varphi^{\mbox{\rm A2}}_j,$ 

\noindent
{\rm (3)}
For each pair of players $\{i,j\} \subseteq N$, 
	if $\varphi_i = \varphi_j$,
	then
	$ \varphi^{\mbox{\rm A2}}_i = \varphi^{\mbox{\rm A2}}_j. $
\end{theorem}

The following theorem provides the number of 
	samples required in Algorithm~A2.

\begin{theorem}\label{sampleA2}
For any $\varepsilon >0$ and  $0< \delta <1$,
	we have the following.

\noindent
{\rm (1)} 
For each player $i \in N=\{1,2,\ldots ,n\}$, 
	if we set 
	$\displaystyle M \geq \frac{\ln 2+ \ln(1/\delta)}{2\varepsilon^2 i^2}$,
	then 
\[
\Prob 
	\left[ 
		\left|
			\varphi^{\mbox{\rm A2}}_i-\varphi_i
		\right| < \varepsilon  
	\right] \geq 1-\delta.
\] 

\noindent
{\rm (2)}
 If we set 	
	$\displaystyle M\geq \frac{\ln 2+\ln(1/\delta)}{2\varepsilon^2}$,
	then 
\begin{eqnarray*}
 \Prob 
	\left[
		\forall i \in N, 
		\left|			
			\varphi^{\mbox{\rm A2}}_i-\varphi_i
		\right| < \varepsilon 
	\right] 
	&\geq& 1-2 \sum_{i=1}^n \left( \frac{\delta}{2} \right)^{i^2} \\
	&=& 1-2
	\left(
					 \left( \frac{\delta}{2} \right)
				 + \left( \frac{\delta}{2} \right)^4
				 + \left( \frac{\delta}{2} \right)^9 + \cdots
				 + \left( \frac{\delta}{2} \right)^{n^2}
		\right).
\end{eqnarray*}

\noindent
{\rm (3)}
 If we set 	
	$\displaystyle M\geq \frac{|N^*| \ln 2+ \ln (1/\delta)}{2\varepsilon^2}$,
	then 
\[
	\Prob
	\left[
		\frac{1}{2}\sum_{i \in N} 
		\left|
			\varphi^{\mbox{\rm A2}}_i-\varphi_i
		\right| < \varepsilon
	\right] \geq 1- \delta,
\]
where $N^*= \{ i \in N\setminus \{n\} 
				\mid \varphi_i > \varphi_{i+1}\} \cup \{n\} $,
	i.e., $|N^*|$ is equal to the size 
	of the maximal player subset, 
	the S-S indices of which are mutually different.
\end{theorem}

\noindent
Proof.  
Let us introduce random variables 
	$X^{(m)}_i \; (\forall m \in \{1,2,\ldots ,M\}, \forall i \in N)$
	in Step~2 of Algorithm~A2 defined by
\[
		X^{(m)}_i=
		\left\{
			\begin{array}{ll}
				1/L^{(m)} & (\mbox{if } 1\leq i \leq L^{(m)}),\\
				0		& (\mbox{if }  L^{(m)}<i).
			\end{array}
		\right.
\]
It is obvious that for each player $i \in N$, 
	$\{ X^{(1)}_i, X^{(2)}_i, \ldots , X^{(M)}_i\}$ 
	is a collection of independent and identically distributed   
	random variables satisfying
	$\varphi^{\mbox{A2}}_i=\sum_{m=1}^M X^{(m)}_i/M$,
 	$ \Exp \left[ \varphi^{\mbox{A2}}_i \right]
	=\Exp \left[ X^{(m)}_i \right]=\varphi_i$, 
	and $1/i \geq X^{(m)}_i \geq 1/n$ 
	$(\forall m \in \{1,2,\ldots ,M\})$.
Hoeffding's inequality~\cite{HOEFFDING1963}
	implies that 
	each player $i \in N$ satisfies
\begin{eqnarray*}
\Prob \left[ \left| \varphi^{\mbox{A2}}_i
	- \varphi_i \right|
	\geq \varepsilon \right]
&\leq 
& 2 \exp \left(
	- \frac{2M^2\varepsilon^2}{\sum_{m=1}^M (1/i-0)^2}
	\right)
= 2 \exp (-2M\varepsilon^2 i^2).
\end{eqnarray*}

\noindent
(1) If we set 
	$\displaystyle M \geq \frac{\ln (2/\delta)}{2\varepsilon^2 i^2}$,
	then 
\[
\Prob \left[ \left| \varphi^{\mbox{A2}}_i
	- \varphi_i \right|
	< \varepsilon \right]\geq 
	1-2 \exp 
	\left(
		-2 \frac{\ln (2/\delta)}{2\varepsilon^2 i^2}
		\varepsilon^2 i^2
	\right)=1-\delta.
\]

\noindent
(2) If we set 
	$\displaystyle M\geq \frac{\ln (2/\delta)}{2\varepsilon^2}$,
	then we have that
\begin{eqnarray*}
\lefteqn{
\Prob \left[
				\forall i \in N, 
					\left| 
						\varphi^{\mbox{A2}}_i	- \varphi_i 
					\right|
					< \varepsilon 
			\right]
 = 1- \Prob 
			\left[
				\exists i \in N, 
					\left| 
						\varphi^{\mbox{A2}}_i	- \varphi_i 
					\right|
					\geq \varepsilon 
			\right]}\\
& \geq 
& 1-\sum_{i \in N} 
	\Prob 
			\left[
					\left| 
						\varphi^{\mbox{A2}}_i	- \varphi_i 
					\right|
					\geq \varepsilon 
			\right]
\geq 1- \sum_{i=1}^n  2 \exp (-2M\varepsilon^2 i^2)\\
& \geq 
& 1- 2 \sum_{i=1}^n   \exp 
	\left( 
		-2  \frac{\ln (2/\delta)}{2\varepsilon^2} 
     \varepsilon^2 i^2
	\right) 
= 1-2\sum_{i=1}^n \left( \frac{\delta}{2} \right)^{i^2}.
\end{eqnarray*}

\noindent
(3) We introduce random variables $Y^{(m)}_{\ell} $
		$(\forall m \in  \{1,2,\ldots ,M\}, \forall \ell \in N)$
	in Step~2 of Algorithm~A2 defined by
\[
		Y^{(m)}_{\ell} = \left\{ \begin{array}{ll}
					1 &  (\mbox{if } \ell=L^{(m)}), \\ 
					0 &  (\mbox{otherwise}).
								\end{array} \right.
\]
Because $\sum_{\ell =1}^n Y^{(m)}_{\ell}=1$ $(\forall m)$, 
	the above definition directly implies that
\[
		X^{(m)}_i=\frac{1}{i}   Y^{(m)}_i
							+\frac{1}{i+1} Y^{(m)}_{i+1}+\cdots 
							+\frac{1}{n}   Y^{(m)}_n.
\]
For each player $i \in N$
	and $i \leq \forall \ell \leq n$, 
	we define 
	$\Pi_{i \ell}=\{ \pi \in \Pi_i \mid {\rm org} (\pi) =\ell\}$.
It is easy to show that
	$|\Pi_{1 \ell}|=|\Pi_{2 \ell}|=\cdots = |\Pi_{\ell \ell}|$
	for each $\ell \in \{1,2,\ldots , n\}$.
The above definitions imply that
\begin{eqnarray*}
\lefteqn{
		\frac{1}{2}\sum_{i \in N} 
		\left|
			\varphi^{\mbox{A2}}_i-\varphi_i
		\right|
= 	\frac{1}{2M}\sum_{i \in N} 
		\left|
			M\varphi^{\mbox{A2}}_i-M\varphi_i
		\right|
=	\frac{1}{2M}\sum_{i \in N} 
		\left|
			\sum_{m=1}^M X^{(m)}_i -M\frac{|\Pi_i|}{n!}
		\right|
} \\
&=& \frac{1}{2M}\sum_{i \in N} 
		\left|
			\sum_{m=1}^M \sum_{\ell=i}^n \frac{1}{\ell}Y^{(m)}_{\ell}
	-\frac{M}{n!} \sum_{\ell=i}^n |\Pi_{i\ell}|
		\right| \\
&=& \frac{1}{2M}\sum_{i \in N} 
		\left|
			\sum_{\ell=i}^n 
			\left(
				\sum_{m=1}^M \frac{1}{\ell}Y^{(m)}_{\ell}
				- \frac{M}{n!}  |\Pi_{i\ell}|
			\right)
		\right| \\
&\leq &  \frac{1}{2M}\sum_{i=1}^n \sum_{\ell=i}^n 
 		\left|
				\sum_{m=1}^M \frac{1}{\ell}Y^{(m)}_{\ell}
				- \frac{M}{n!}  |\Pi_{i\ell}|
		\right| 
= \frac{1}{2M}\sum_{\ell=1}^n \sum_{i=1}^\ell 
 		\left|
				\sum_{m=1}^M \frac{1}{\ell}Y^{(m)}_{\ell}
				- \frac{M}{n!}  |\Pi_{i\ell}|
		\right|  \\
&=&  \frac{1}{2M}\sum_{\ell=1}^n \ell
 		\left|
				\sum_{m=1}^M \frac{1}{\ell}Y^{(m)}_{\ell}
				- \frac{M}{n!}  |\Pi_{1\ell}|
		\right| 
 \;\;\;(\mbox{since }\; |\Pi_{1 \ell}|=|\Pi_{2 \ell}|
		=\cdots = |\Pi_{\ell \ell}|) \\
&=&  \frac{1}{2M}\sum_{\ell=1}^n
 		\left|
				\sum_{m=1}^M Y^{(m)}_{\ell}
				- \frac{M \ell}{n!} |\Pi_{1\ell}|
		\right|. 
\end{eqnarray*}

For each player $\ell \not \in N^*$, 
	we have the equalities $|\Pi_{\ell}|=n! \varphi_{\ell}
			=n! \varphi_{\ell+1}=|\Pi_{\ell+1}|$, 
	which yields that 
	$f_{\ell+1}:\Pi_{\ell+1} \rightarrow \Pi_{\ell}$
	is a bijection and thus $\Pi_{\ell}$ does not include
	any originator.
From the above, it is obvious that, 
	if   $\ell \not \in N^*$,  then 
	\mbox{$\Pi_{1\ell}=\Pi_{2\ell}=\cdots =\Pi_{\ell \ell}=\emptyset$}.
 For each $\ell \in \{1,2,\ldots ,n\}$,
	$\{Y^{(1)}_{\ell}, Y^{(2)}_{\ell},\ldots , Y^{(M)}_{\ell}\}$
	is a Bernoulli process satisfying
  $\Exp [Y^{(m)}_\ell]
		=\frac{1}{n!}\sum_{i=1}^{\ell}|\Pi_{i \ell}|
		=\frac{\ell}{n!} |\Pi_{1 \ell}|$ $(\forall m)$.
Thus, 	 $\ell \not \in N^*$ implies that
	$Y^{(m)}_\ell=0$ for any $m \in \{1,2,\ldots ,M\}$.
To summarize the above, 
	we have shown that
\[
	 \mbox{ if } \ell \not \in N^*
	 \mbox{ then } 
				\sum_{m=1}^M Y^{(m)}_{\ell}
				- \frac{M \ell}{n!}|\Pi_{1\ell}|
	 =\sum_{m=1}^M 0-\frac{M \ell}{n!}0=0. 
\] 

Now, we have an upper bound of the total variation distance
\begin{eqnarray*}
		\frac{1}{2}\sum_{i \in N} 
		\left|
			\varphi^{\mbox{A2}}_i-\varphi_i
		\right|
&\leq&  \frac{1}{2M}\sum_{\ell=1}^n
 		\left|
				\sum_{m=1}^M Y^{(m)}_{\ell}
				- \frac{M \ell}{n!} |\Pi_{1\ell}|
		\right| \\
&=& \frac{1}{2M}\sum_{\ell \in N^*}
 		\left|
				\sum_{m=1}^M Y^{(m)}_{\ell}
				- \sum_{m=1}^M \Exp [Y^{(m)}_{\ell}]
		\right|. 
\end{eqnarray*}

Obviously, the vector of random variables
$
	 \left(
		\sum_{m=1}^M Y^{(m)}_{\ell}
		\right)_{\ell \in N^*}
$
	is multinomially distributed and satisfies that
	the total sum is equal to $M$.
Then, the Bretagnolle-Huber-Carol inequality~\cite{VAART1989}
	(Theorem~\ref{BHC} in Appendix)
	implies that
\begin{eqnarray*}
\lefteqn{
	\Prob
	\left[
		\frac{1}{2}\sum_{i \in N} 
		\left|
			\varphi^{\mbox{A2}}_i-\varphi_i
		\right| \geq \varepsilon
	\right]
 \leq 
	\Prob
	\left[
		\frac{1}{2M}\sum_{\ell \in N^*}
 		\left|
				\sum_{m=1}^M Y^{(m)}_{\ell}
				- \sum_{m=1}^M \Exp [Y^{(m)}_{\ell}]
		\right| \geq \varepsilon
	\right] 
}\\
&=& 
	\Prob
	\left[
		\sum_{\ell \in N^*}
 		\left|
				\sum_{m=1}^M Y^{(m)}_{\ell}
				- \sum_{m=1}^M \Exp [Y^{(m)}_{\ell}]
		\right| \geq 2M \varepsilon
	\right] 
\leq 2^{|N^*|} \exp 
	\left( -2M \varepsilon^2 \right) \\
&\leq & 2^{|N^*|} \exp 
	\left( 
		-2 
			\frac{\ln \left( 2^{|N^*|}/\delta \right)}{2\varepsilon^2}
 		\varepsilon^2 
	\right) = \delta \\
\end{eqnarray*}
 and thus, we have the desired result.
\qed

\smallskip

The following corollary provides an approximate version
	of Theorem~\ref{sampleA2}~(2).
Surprisingly, it says that
	the required number of samples is irrelevant 
	to $n$ (number of players). 

\begin{corollary}
For any $\varepsilon >0$ and  $0< \delta' <1$,
	we have the following.
 If we set 	
	\mbox{
		$\displaystyle 
		M\geq \frac{\ln 2+ \ln(1/\delta')+\ln 1.129}{2\varepsilon^2}$,
	}
	then 
\[
 \Prob 
	\left[
		\forall i \in N, 
		\left|			
			\varphi^{\mbox{\rm A2}}_i-\varphi_i
		\right| < \varepsilon 
	\right] \geq 1-\delta'.
\]
\end{corollary}

\noindent
Proof. 
If we put $\delta = \delta'/1.129$,
	then Theorem~\ref{sampleA1}~(2) implies that
\begin{eqnarray*}
\lefteqn{\Prob 
	\left[
		\forall i \in N, 
		\left|			
			\varphi^{\mbox{A2}}_i-\varphi_i
		\right| < \varepsilon 
	\right] 
}\\
	&\geq & 1- 2
	\left(
					 \left( \frac{\delta}{2} \right)
				 + \left( \frac{\delta}{2} \right)^4
				 + \left( \frac{\delta}{2} \right)^9 + \cdots
				 + \left( \frac{\delta}{2} \right)^{n^2}
		\right) \\
&\geq &
1-\delta 
	\left(1
		+ \left( \frac{1}{2} \right)^3
		+ \left( \frac{1}{2} \right)^8
		+ \left( \frac{1}{2} \right)^{15} 
		+ \left( \frac{1}{2} \right)^{24} + \cdots
		+ \left( \frac{1}{2} \right)^{n^2-1}
	\right)		\\
&\geq &
1-\delta 
	\left(1
		+ \left( \frac{1}{2} \right)^3
		+ \left( \frac{1}{2} \right)^8
			\left(1
				+ \left( \frac{1}{2} \right)^{7}
				+ \left( \frac{1}{2} \right)^{14}
				+ \left( \frac{1}{2} \right)^{21}+ \cdots
			\right)
	\right)		\\
&= & 1- \delta 
		\left(1+ 
			\left(
				\frac{1}{2}
			\right)^3 +
			\left(
				\frac{1}{2}
			\right)^8
			\left( 
				\frac{1}{1-(1/2)^7}
			\right)
		\right)
		\geq  1-1.129 \delta =1-\delta'.  \\[-7ex]
\end{eqnarray*}
\qed

\smallskip

\noindent
Here, we note that 
$\ln 2     \simeq 0.69314$ and 
$\ln 1.129 \simeq 0.12133$. 

In a practical setting, it is difficult 
	to estimate the size of $N^*$ 
	defined in  Theorem~\ref{sampleA2}~(3), 
	since the problem of verifying the asymmetricity
	of a given pair of players is 
	\mbox{NP-complete}~\cite{MATSUI2001}.
The following corollary is useful in some practical situations.

\begin{corollary}
For any $\varepsilon >0$ and  $0< \delta <1$,
	we have the following.
 If we set 	
	\mbox{
		$\displaystyle 
		M\geq \frac{n'' \ln 2+ \ln (1/\delta)}{2\varepsilon^2}$,
	}
	then 
\[
	\Prob
	\left[
		\frac{1}{2}\sum_{i \in N} 
		\left|
			\varphi^{\mbox{\rm A2}}_i-\varphi_i
		\right| < \varepsilon
	\right] \geq 1- \delta,
\]
where $n''=| \{ i \in N\setminus \{n\} 
				\mid w_i > w_{i+1}\} \cup \{n\} |$,
	i.e., $n''$ is equal to the size of a maximal player subset 
	with mutually different weights.
\end{corollary}

\noindent
Proof. 
Since $\varphi_i > \varphi_{i+1}$ implies $w_i > w_{i+1}$,
	it is obvious that $|N^*| \leq n''$ 
	and we have the desired result. 
\qed

\smallskip

A game of the power of the countries in the EU Council 
	is defined by
\[
	[255; 29, 29, 29, 29, 27, 27, 14, 13, 12, 12, 12, 12, 12, 10, 10, 10, 7, 7, 7, 7, 7, 4, 4, 4, 4, 4, 3] 
\]~\cite{FELSENTHAL2001,BILBAO2002}.
In this case, $n=27$ and $n''=9$.
A weighted majority game defined by~\cite{OWEN1995}(Section~12.4)
	for a voting process in United States
	has a vector of weights
\[
\begin{array}{l}
[270; 
	45, 
	41, 
	27, 
	26, 26, 
	25, 
	21, 
	17, 17, 
	14, 
	13, 13, 
	12, 12, 12, 
	11,
	\underbrace{10, \ldots , 10}_{\mbox{4 times}},
	\underbrace{9, \ldots , 9}_{\mbox{4 times}},
	\\
	8,  8,  
	\underbrace{7, \ldots , 7}_{\mbox{4 times}},
	\underbrace{6, \ldots , 6}_{\mbox{4 times}},
	5,
	\underbrace{4, \ldots , 4}_{\mbox{9 times}},
	\underbrace{3, \ldots , 3}_{\mbox{7 times}}
],
\noindent
\mbox{ where } n=51 \mbox{ and } n''=19.
\end{array}
\]

\section{Computational Experiments}

This section reports the results of our preliminary numerical experiments.
All the experiments were conducted on a windows machine,
i7-7700 CPU@3.6GHz Memory (RAM) 16GB.
Algorithms A1 and A2 are implemented by Python 3.6.5. 

\begin{figure}[ht]
\begin{center}
\includegraphics[width=18cm, bb=0 0 958 434]{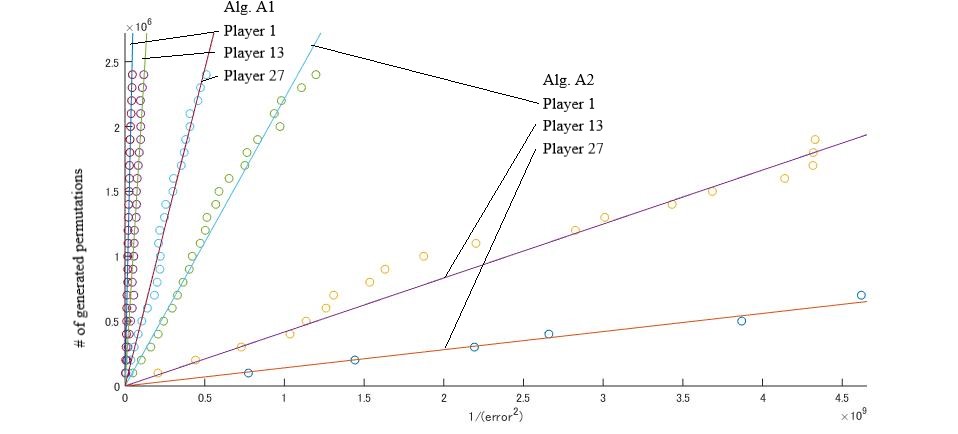}
\caption{EU Council.} \label{EU}
\end{center}
\end{figure}

\begin{figure}[ht]
\begin{center}
\includegraphics[width=18cm, bb=0 0 958 434]{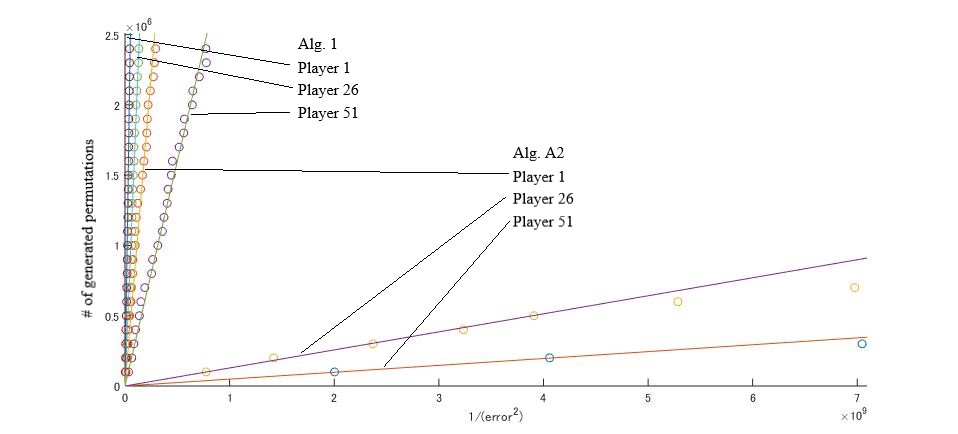}
\caption{United States.} \label{US}
\end{center}
\end{figure}

We tested the EU Council instance and the United States instance
described in the previous section.
In each instance,  we set  $M$ in Algorithm~A1 and A2 
	(the number of generated permutations) to 
	$M \in \{1 \times 10^5, 2 \times 10^5, \ldots , 24 \times 10^5\}$.
For each value $M$, we executed Algorithms~A1 and A2, 100 times.
Figures~\ref{EU} and~\ref{US} show results of some players.
For each value $M$, 
	we calculated  the mean number of  $|\varphi_i -\varphi^{\mbox{A}}_i |$,
	denoted by $\widehat{\varepsilon_i}$,
	in an average of 100 trials. 
The horizontal axes of  Figures~\ref{EU} and~\ref{US} show the value $1/\widehat{\varepsilon_i}^2$.
Under the assumption that $M=\alpha /\widehat{\varepsilon_i}^2$, 
	we estimated $\alpha$ by the least squares method.
Table~\ref{alpha} shows the results and ratios of $\alpha$ of two algorithms. 

\begin{table}[ht]
\begin{center}
\caption{Comparison of Algorithms A1 and A2.} \label{alpha}
\begin{tabular}{|l|r|r|r|}
\hline
EU Council &\multicolumn{2}{c|}{$\alpha $} & \\
\cline{2-3}
	& Alg. A1 & Alg. A2 & ratio \\
\hline
Player 1 & 0.0557 & 0.0022 & 25.318 \\
Player 13 & 0.0199 & $4.1615 \times 10^{-4} $& 47.819 \\
Player 27 & 0.0049 & $1.3987 \times 10^{-4}$ & 35.033 \\
\hline
\end{tabular}

\medskip

\begin{tabular}{|l|r|r|r|}
\hline
United States &\multicolumn{2}{c|}{$\alpha $} & \\
\cline{2-3}
	& Alg. A1 & Alg. A2 & ratio \\
\hline
Player 1 & 0.0489 & 0.0181 & 2.7017 \\
Player 26 & 0.0088 & $1.2837 \times 10^{-4} $& 68.552 \\
Player 51 & 0.0032 & $4.8911 \times 10^{-5}$ & 65.424 \\
\hline
\end{tabular}
\end{center}
\end{table}

\begin{figure}[ht]
\begin{center}
\includegraphics[width=18cm, bb=0 0 958 434]{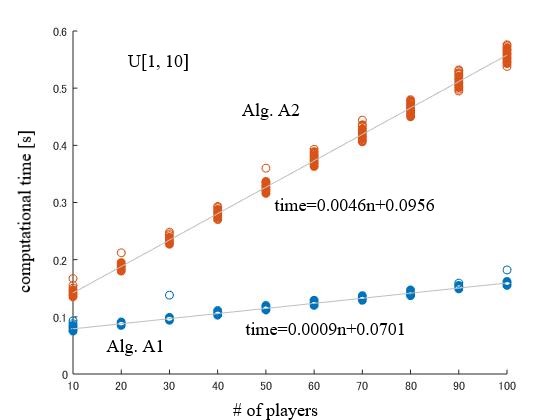} 
\includegraphics[width=18cm, bb=0 0 958 434]{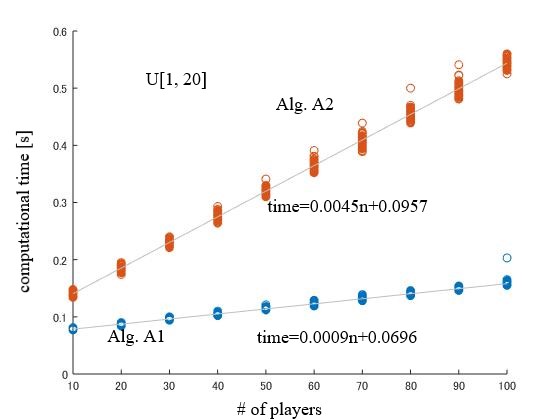}
\caption{Computational time.} \label{CompTime}
\end{center}
\end{figure}

For each (generated) permutation, the computational effort 
	of both Algorithms A1 and A2 are bounded by $\Order (n)$.
Here, we discuss the constant factors of $\Order (n)$ computations.
We tested the cases that weights $w_i$ are generated uniformly at random from the intervals 
	$[1, 10]$ or $[1, 20]$, and quota is equal to $(1/2)\sum_{i \in N}w_i$.
For each $n \in \{10, 20, \ldots , 100\}$, 
	we executed Algorithms A1 and A2 by setting $M=10,000$.
Under the assumption that computational time is equal to $an+b$, 
	we estimated $a$ and $b$ by the least squares method.
Figure~\ref{CompTime} shows that
	for each permutation, the computational effort of Algorithm A2 increases 
	about 5-fold comparing to Algorithm A1.  

\section{Conclusion}

In this paper,
	we analyzed a naive Monte Carlo algorithm
	(Algorithm~A1) for calculating the S-S index
	denoted by 	$(\varphi_1,\varphi_2,\ldots ,\varphi_n)$
 	in weighted majority games.
By employing the Bretagnolle-Huber-Carol
	inequality~\cite{VAART1989}
	(Theorem~\ref{BHC} in Appendix), 
	we estimated the required number of samples 
	that gives an upper bound of the total variation distance.

We also proposed an efficient	Monte Carlo algorithm
	(Algorithm~A2).
The time complexity of each iteration of our algorithm is equal to 
	that of the naive algorithm (Algorithm~A1). 
Our algorithm has the property that
	the obtained estimator 
	$(\varphi^{\mbox{A2}}_1,\varphi^{\mbox{A2}}_2,
		\ldots ,\varphi^{\mbox{A2}}_n)$
	satisfies
\[
\mbox{ both }
[
	\mbox{ if } \varphi_i < \varphi_j
	\mbox{ then }
	\varphi^{\mbox{A2}}_i \leq \varphi^{\mbox{A2}}_j
]
\mbox{ and }
[
	\mbox{ if } \varphi_i = \varphi_j
	\mbox{ then }
	\varphi^{\mbox{A2}}_i = \varphi^{\mbox{A2}}_j
].
\]
We also proved that, 
	even if we consider the property
\[
 \Prob 
	\left[
		\forall i \in N, 
		\left|			
			\varphi^{\mbox{A2}}_i-\varphi_i
		\right| < \varepsilon 
	\right] \geq 1-\delta,
\]
	the required number of samples 
	is irrelevant to $n$ (the number of players).


\section*{APPENDIX (Bretagnolle-Huber-Carol inequality)}
 
\begin{theorem}\label{BHC}
	\mbox{\rm \cite{VAART1989} }

If the random vector $(Z_1, Z_2, \ldots ,Z_n)$
	is multinomially distributed with parameters 
	$(p_1,p_2,\ldots ,p_n)$ and satisfies 
	$Z_1+Z_2+ \cdots +Z_n=M$
	then 
\[
	\Prob 
	\left[
		\sum_{i=1}^n
		\left|
			Z_i-Mp_i
		\right|
		\geq 2M \varepsilon
	\right]
	\leq 2^n \exp (-2M \varepsilon^2).
\]
\end{theorem}

\noindent
Proof. 
It is easy to see that
\begin{eqnarray*}
\lefteqn{
	\Prob 
	\left[
				\sum_{i=1}^n
		\left|
			Z_i-Mp_i
		\right|
		\geq 2M \varepsilon
	\right]
=	\Prob 
	\left[
		2 \max_{S \subseteq \{1,2,\ldots ,n\}}
			\sum_{i \in S} ( Z_i -Mp_i)
		\geq 2M \varepsilon
	\right]
}\\
&=&\Prob 
	\left[
		\exists S \subseteq \{1,2,\ldots ,n\}, \;\;
			\sum_{i \in S} ( Z_i -Mp_i)
		\geq M \varepsilon
	\right]
\leq \sum_{S \subseteq \{1,2,\ldots ,n\}}
	\Prob
	\left[
			\sum_{i \in S} ( Z_i -Mp_i)
		\geq M \varepsilon
	\right]\\
&=& \sum_{S \subseteq \{1,2,\ldots ,n\}}
	\Prob
	\left[
			\sum_{i \in S} Z_i 
			-M\sum_{i \in S} p_i
		\geq M \varepsilon
	\right]\\
\end{eqnarray*}

For any subset $S \subseteq \{1,2,\ldots ,n\}$,
	there exists a Bernoulli process
	$(X^{(1)}_S, X^{(2)}_S, \ldots , X^{(M)}_S)$
	satisfying $\sum_{i \in S} Z_i=\sum_{m=1}^M X^{(m)}_S$
	and $\Exp [X^{(m)}_S]=\sum_{i \in S} p_i$ 
	$(\forall m \in \{1,2,\ldots M\})$.
Hoeffding's inequality~\cite{HOEFFDING1963}
	implies that 
\begin{eqnarray*}
\lefteqn{
\sum_{S \subseteq \{1,2,\ldots ,n\}}
	\Prob
	\left[
			\sum_{i \in S} Z_i 
			-M\sum_{i \in S} p_i
		\geq M \varepsilon
	\right]
=\sum_{S \subseteq \{1,2,\ldots ,n\}}
	\Prob
	\left[
			\sum_{m=1}^M X^{(m)}_S
			-\Exp 
			\left[
				\sum_{m=1}^M X^{(m)}_S
			\right]
		\geq M \varepsilon
	\right]
}\\
&=& \sum_{S \subseteq \{1,2,\ldots ,n\}}
	\Prob
	\left[
			\frac{1}{M}\sum_{m=1}^M X^{(m)}_S
			-\frac{1}{M}\Exp 
			\left[
				\sum_{m=1}^M X^{(m)}_S
			\right]
		\geq  \varepsilon
	\right]
\leq  \sum_{S \subseteq \{1,2,\ldots ,n\}}
	\exp (-2M\varepsilon^2)\\
&=& 2^n \exp (-2M\varepsilon^2). \\[-7ex]
\end{eqnarray*}
\QED


\begin{thebibliography}{}

\bibitem
{BACHRACH2010}
Bachrach,~Y., Markakis,~E., Resnick,~E., Procaccia,~A.~D., Rosenschein,~J.~S., and Saberi,~A. 
\newblock Approximating power indices: theoretical and empirical analysis. 
\newblock {\em Autonomous Agents and Multi-Agent Systems}, 
	{\bf 20}~(2010), 105--122.

\bibitem
{BERGHAMMER2011}
Berghammer,~R., Bolus,~S., Rusinowska,~A., and De Swart,~H. 
\newblock A relation-algebraic approach to simple games. 
\newblock {\em European Journal of Operational Research}, 
	{\bf 210}~(2011), 68--80.

\bibitem
{BILBAO2000}
Bilbao,~J.~M., Fernandez,~J.~R., Losada,~A.~J., and Lopez,~J.~J.
\newblock Generating functions for computing power indices efficiently. 
\newblock {\em Top}, 
	{\bf 8}~(2000), 191--213.

\bibitem
{BILBAO2002}	
Bilbao,~J.~M., Fernandez,~J.~R., Jim\'{e}nez,~N., and Lopez,~J.~J. 
\newblock Voting power in the European Union enlargement.
\newblock {\em European Journal of Operational Research}, 
  {\bf 143}~(2002),  181--196.	

\bibitem
{BOLUS2011}
Bolus,~S.
\newblock Power indices of simple games and vector-weighted majority games by means of binary decision diagrams. 
\newblock {\em European Journal of Operational Research}, 
	{\bf 210}~(2011), 258--272.

\bibitem
{BRAMS1975}
Brams,~S.~J. and P. J. Affuso,~P.~J.
\newblock Power and size; a new paradox. 
\newblock {\em Mimeographed Paper}, 1975.

\bibitem
{CASTRO2009}
Castro,~J., G\'omez,~D., and Tejada,~J.
\newblock Polynomial calculation of the Shapley value based on sampling.
\newblock {\em Computers \& Operations Research}, 
	{\bf 36} (2009), 1726--1730.

\bibitem
{CASTRO2017}
Castro,~J., G\'omez,~D., Molina,~E., and Tejada,~J.
\newblock Improving polynomial estimation of the Shapley value by stratified random sampling with optimum allocation. 
\newblock {\em Computers \& Operations Research}, 
	{\bf 82} (2017), 180--188.

\bibitem
{DENG1994}
Deng,~X. and Papadimitriou,~C.~H.
\newblock On the complexity of cooperative solution concepts. 
\newblock {\em Mathematics of Operations Research}, 
	{\bf 19}~(1994), 257--266.

\bibitem
{ELKIND2007}
Elkind,~E., Goldberg,~L.~A., Goldberg,~P., and Wooldridge,~M.
\newblock Computational complexity of weighted threshold games. 
\newblock In \emph{Proc. of the National Conference on Artificial Intelligence},
	AAAI Press, 718--723, 2007.

\bibitem
{FATIMA2008}
Fatima,~S.~S., Wooldridge,~M., and Jennings,~N.~R. 
\newblock A linear approximation method for the Shapley value
\newblock {\em Artificial Intelligence}, 
	{\bf 172} (2008), 1673--1699.

\bibitem
{FELSENTHAL2001}
Felsenthal,~D.~S. and Machover,~M.  
\newblock The Treaty of Nice and qualified majority voting. 
\newblock {\em Social Choice and Welfare},
  {\bf 18} (2001), 431--464.


\bibitem
{GAREY1979}
Garey,~M.~R. and Johnson,~D.~S.
\newblock Computers and Intractability: A Guide to the Theory of NP-Completeness.
\newblock  WH Freeman, 1979.

\bibitem
{HOEFFDING1963}
Hoeffding,~W.
\newblock Probability inequalities for sums of bounded random variables,
\newblock {\em Journal of the American Statistical Association}, 
	{\bf 58}~(1963), 13--30.

\bibitem
{KLINZ2005}
Klinz,~B. and Woeginger,~G.~J.
\newblock Faster algorithms for computing power indices in weighted voting games.
\newblock {\em Mathematical Social Sciences}, 
	{\bf 49}~(2005), 111--116.

\bibitem
{LEECH2003}
Leech,~D.
\newblock Computing power indices for large voting games. 
\newblock {\em Management Science}, 
	{\bf 49}~(2003), 831--837.

\bibitem
{LUCAS1983}
Lucas,~W.~F.
\newblock Measuring power in weighted voting systems. 
\newblock Brams,~S.~J., Lucas,~W.~F., and Straffin,~P.~D. (eds.): 
	Political and Related Models, Springer, 183--238, 1983.

\bibitem
{MANN1960}
Mann,~I. and Shapley,~L.~S. 
\newblock Values of large games. IV: evaluating the electoral college by Montecarlo techniques.
\newblock {\em Technical Report, The RAND Corporation}, 
{\bf RM-2651}, 1960.

\bibitem
{MANN1962}
Mann,~I. and Shapley,~L.~S. 
\newblock Values of large games. VI: Evaluating the electoral college exactly.
\newblock {\em Technical Report, The RAND Corporation}, 
{\bf RM-3158-PR}, 1962.

\bibitem
{MATSUI2000}
Matsui,~T. and Matsui,~Y. 
\newblock A survey of algorithms for calculating power indices of weighted majority games. 
\newblock {\em Journal of the Operations Research Society of Japan},
	{\bf 43}~(2000), 71--86.

\bibitem
{MATSUI2001}
Matsui,~Y. and Matsui,~T.
\newblock NP-completeness for calculating power indices of weighted majority games.
\newblock {\em Theoretical Computer Science}, 
	{\bf 263}~(2001), 305--310.

\bibitem
{OWEN1972}
Owen,~G. 
\newblock Multilinear extensions of games. 
\newblock {\em Management Science}, 
	{\bf 18}~(1972), 64--79.

\bibitem
{OWEN1995}
Owen,~G.
\newblock Game Theory.
\newblock Academic press, 1995.

\bibitem
{PRASAD1990}
Prasad,~K. and Kelly,~J.~S.
\newblock NP-completeness of some problems concerning voting games. 
\newblock {\em International Journal of Game Theory}, 
	{\bf 19}~(1990), 1--9.

\bibitem
{RUSHDI2017}
Rushdi,~A.~M.~A. and Ba-Rukab,~O.~M.
\newblock Map calculation of the Shapley-Shubik voting powers: An example of the European Economic Community.
\newblock {\em International Journal of Mathematical, Engineering and Management Sciences (IJMEMS)}, 
	{\bf 2}~(2017), 17--29.

\bibitem
{SHAPLEY1953}
Shapley,~L.~S.
\newblock A value for $n$-person games.
\newblock Kuhn,~H.~W. and Tucker,~A.~W. Tucker (eds.),
 Contributions to the Theory of Games II,
 Princeton University Press, 307--317, 1953.

\bibitem
{SHAPLEY1954}
Shapley,~L.~S. and Shubik,~M. 
\newblock An algorithm for evaluating the distribution of power in a committee system. 
\newblock {\em American Political Science Review}, 
	{\bf 48}~(1954), 787--792.


\bibitem
{UNO2012}
Uno,~T. 
\newblock Efficient computation of power indices for weighted majority games.
In \emph{Proc. of International Symposium on Algorithms and Computation (ISAAC)}, 
LNCS {\bf 7676}, Springer, 679--689, 2012.


\bibitem
{VAART1989}
van der Vaart,~A.~W. and Wellner,~J.~A.
\newblock Weak Convergence and Empirical Processes: with Applications to Statistics, 
\newblock Springer, 1989.

\bibitem
{NEUMANN1944}
von~Neumann,~J. and Morgenstern,~O.
\newblock Theory of Games and Economic Behavior.
\newblock Princeton Univ. Press, 1944.

\end{thebibliography}
\end{document}